\documentstyle[buckow1,12pt]{article}

\def\kreiso{\lower0.85pt\hbox{\Large $\bullet$}}
\def\kreisv{\raise0.85pt\hbox{$\scriptstyle\bigcirc$}}
\def\bbox{\lower0.85pt\hbox{$\Box$}\,}

\newcommand{\includemygpt}[2]{%
\centering%
  {\fontsize{15pt}{18pt}\selectfont%
      \input{#1.tex}}%
\caption[]{#2}%
\label{fig:#1}%
  }

\begin {document}

\vspace{-3.5cm}
\begin{flushleft}
{\normalsize DESY 98-043} \hfill\\
{\normalsize April 1998} 
\end{flushleft}

\def\titleline{
Lattice Chiral Schwinger Model: Selected Results\footnote{Talk given
  by V. Bornyakov at 31$^{\mbox{st}}$ International Ahrenshoop Symposium
  on the Theory of Elementary Particles, Buckow, September 1997.}
}
\def\authors{
V. Bornyakov \1ad, G. Schierholz \2ad \3ad and A. Thimm \2ad \4ad
}
\def\addresses{
\1ad 
        Institute for High Energy Physics IHEP,
        RU-142284 Protvino, Russia \\
\2ad        
        Deutsches Elektronen-Synchrotron DESY and HLRZ, \\
        D-15735 Zeuthen, Germany \\

\3ad        
        Deutsches Elektronen-Synchrotron DESY, D-22603 Hamburg, Germany \\

\4ad 
       Institut f\"ur Theoretische Physik, Freie
        Universit\"at Berlin, \\
        D-14195 Berlin, Germany
}       


\def\abstracttext{
We discuss a method for regularizing chiral gauge
theories. The idea is to formulate the gauge
fields on the lattice, while the fermion determinant
is regularized and computed in the continuum. A simple effective action
emerges which lends itself to numerical simulations.
}

\makefront

\section{Introduction}

The regularization of chiral fermions by means of a space-time lattice
has well-known difficulties. To evade them it has been 
suggested~\cite{AGDP,GS1,tHooft} to discretize only the gauge fields 
and treat the fermions in the continuum by introducing an interpolation 
of the latter~\cite{GKSW}. In the present talk we shall test the idea
in the chiral Schwinger model. First results of this approach 
have been reported in~\cite{BST}. For similar ideas see~\cite{Bodwin}.

To compute the effective action we start from Wilson fermions. The
action is
\begin{eqnarray}
 S_{\pm} & =  &  \frac{1}{2a_f}\sum_{n,\mu} \left\{\bar{\psi}(n) \gamma_{\mu}
 [(P_{\mp}+P_{\pm} U^f_{\mu}(n))\psi(n+\mu)\right. \nonumber \\
 & - & \left.(P_{\mp}+P_{\pm} U^{f \,\dagger}_{\mu}(n-\mu))\psi(n-\mu)]\right\} \\
         & +  & S_W(U^f), \label{action}
\end{eqnarray}
where $P_\pm = (1 \pm \gamma_5)/2$, and where we have denoted the
lattice spacing of the fermionic lattice by $a_f$. Later on we will
take the limit $a_f \rightarrow 0$. In practice $a_f = a/N$, $N$
integer, where $a$ is the lattice spacing of the gauge field action. 
The link variables on the fine lattice, $U^f$, are obtained by a suitable
interpolation~\cite{GKSW} from the links on the original lattice, $U$.
The usual (gauged) Wilson term $S_W$ reads
\begin{equation}
S_W(U^f) = \frac{1}{2a_f}\sum_{n,\mu} \bar{\psi}(n) 
[2\psi(x)- U^f_{\mu}(n)\psi(n+\mu) - U^{f \,\dagger}_{\mu}(n-\mu)
\psi(n-\mu)]. 
\end{equation}
We will consider ungauged, $S_W(U^f=1)$, and partially
gauged Wilson terms as well.
 
The effective action is obtained in three steps. First one defines
\begin{equation}
\exp(-W_\pm) = \int {\cal D}\psi {\cal D}\bar{\psi}
	 \exp(-S_\pm).
\end{equation}
Then one performs the limit $\lim_{a_f \rightarrow 0} W_\pm$. One 
finds that this action is not invariant under chiral gauge
transformations, not even in the anomaly-free model.
Gauge invariance can, however, be restored by adding a local 
counterterm.\footnote{It was shown in \cite{tHooft} for a rather
general class of gauge fields, not including compact $U(1)$ fields
though, that the
effective action can be made gauge invariant and finite.} The
counterterm can be identified analytically and its
coefficient be calculated in one-loop perturbation theory~\cite{BST2}. It is
\begin{equation}
c \sum_x A_\mu^2 (x),
\end{equation}
where $A_\mu = \lim_{a_f \rightarrow 0}\, (1/a_f)\, \mbox{Im}\, U^f_\mu$, and
$c = - 0.0202$\footnote{This value can also be extracted from~\cite{SZ}.}
for both gauged and ungauged Wilson terms.
We then arrive at the effective action
\begin{equation}
\widehat{W}_\pm = \lim_{a_f \rightarrow 0} W_\pm 
+ c \sum_x A_\mu^2(x).
\end{equation}

The effective action $W_\pm$ has been computed by means of the Lanczos
method. Note
that the action is non-hermitean. The
Lanczos vectors were re-orthogonalized after every iteration.
For smooth gauge fields it was shown that~\cite{BST}
\begin{equation}
\mbox{Re}\widehat{W}_\pm = \frac{1}{2}(W + W_0),
\end{equation}
where $W$ ($W_0$) is the effective action of the vector model (free
theory). This result was conjectured in~\cite{AGDP} for `perturbative'
gauge fields. 

Because of lack of space we shall restrict ourselves in the written
version of the talk to rough gauge fields. We assume that the reader
is familiar with the problem and with previous results on the subject.  

\section{Effective action revisited}

Before we turn to the problem of rough gauge fields, let us briefly mention 
some new results on the effective action. 

On the $L^2$ torus the gauge field can be written 
\begin{equation}
 A_{\mu}(x) = \frac{2\pi}{L} t_{\mu} + \varepsilon_{\mu\nu} \partial_{\nu}
 \alpha(x) + \mbox{i} g^{-1}(x) \partial_{\mu} g(x),\; 
g(x) = \exp(-\mbox{i}\beta(x))
\label{field}
\end{equation}
assuming $A_\mu(x) \in [-\pi,\pi)$, where
$t_\mu \in [-1,1)$ are the zero momentum modes (torons), 
$\bbox \alpha(x) = - F_{12}(x)$ and $g(x) \in U(1)$ is a gauge
transformation.\footnote{As we shall see this assignment
is not unique if we allow large gauge transformations.} We assume 
periodic boundary conditions
for the gauge fields and antiperiodic
boundary conditions for the fermions.

Writing $a_\mu = (2\pi/L)\, t_\mu$ and $\widetilde{A}_\mu = A_\mu -
a_\mu$, the real part of the effective action factorizes in the form
\begin{equation}
\mbox{Re} \widehat{W}_\pm(A) = \mbox{Re}\widehat{W}_\pm(\widetilde{A}) +  
\mbox{Re} \widehat{W}_\pm(a) + W_0.
\label{fac}
\end{equation}
Formally this relation follows from the property of the Dirac operator 
\begin{equation}
\not{\!\!D}(\widetilde{A} + a)\exp(-\mbox{i}\beta + \gamma_5\alpha) = 
\exp(-\mbox{i}\beta - \gamma_5\alpha) \not{\!\!D}(a).
\end{equation}
A similar expression can be derived for the imaginary part. The toron
part of the effective action, $\widehat{W}_\pm(a)$, is known
analytically~\cite{GS2}. For the fluctuating part we obtain
\begin{equation}
\widehat{W}_\pm(\widetilde{A})  = \sum_{\alpha}  \frac{q^2_\alpha}{8\pi}
\int \mbox{d}^2 x \,\widetilde{A}_\mu \left[\delta_{\mu\nu} - 
(\partial_\mu + \mbox{i}\epsilon_\alpha \widetilde{\partial}_\mu) 
\frac{1}{\Box} (\partial_\nu + \mbox{i}\epsilon_\alpha 
\widetilde{\partial}_\nu)\right]\widetilde{A}_\nu,
\label{r2}
\end{equation}
where $\widetilde{\partial}_\mu = \varepsilon_{\mu\nu} \partial_\nu$, and 
$q_\alpha$ and $\epsilon_{\alpha}$ are the fermion charge and
chirality, respectively. This result is in agreement with the well
known expression for the effective action in 
${\sf R}\!\!\!{\sf R}^2$~\cite{JR}. Note that the imaginary part of
(\ref{r2}) vanishes in the anomaly-free model.

To check the result (\ref{r2}) and our method of calculation, 
we discretized the continuum 
configuration~\cite{NN3} $A_\mu(x) = c_\mu \cos(2\pi kx/L) + (2\pi/L)\,t_\mu$ 
with $c_1 = c_2 = 0.32$ and $k_1 = 1$, $k_2 = 0$ (as chosen
in~\cite{NN3}) and computed $W_\pm$.
In Fig.~1 we show $\mbox{Im}\, {W}_\pm (A)$ as a function of $r =
a_f/a$ for two toron fields.
For the extrapolated values we obtain $- 0.004078$ and $0.002098$,
respectively. This is to be compared with the analytical values $-0.004074$ 
and $0.002094$, respectively. We find the same good agreement with the
analytic formulae for the real part.

In~\cite{BST2} we shall give an analytic
proof of eqs.~(\ref{fac}) and (\ref{r2}).

\section{Vortex-antivortex configuration}

In \cite{BST} we reported numerical evidence for gauge invariance of 
the anomaly-free effective action under a class of gauge transformations. 
This did not include `singular' gauge transformations which create a
vortex-antivortex pair.\footnote{Note that singular gauge transformations
generally create a problem in the overlap approach \cite{NN1}.}

Let us consider a lattice gauge field configuration
\begin{equation}
\theta_{\mu}(s) = \theta^v_{\mu}(s-v) - \theta^v_{\mu}(s-\bar{v}) -
     \frac{2\pi}{L^2} \varepsilon_{\mu\nu} (v_\nu-\bar{v}_\nu),  
\end{equation}
where $\theta^v_{\mu}(s) = 2\pi \varepsilon_{\mu\nu} \partial_\nu G(s)$, 
$G(s)$ being the lattice inverse Laplacian. This configuration
corresponds to a vortex-antivortex pair at positions $v$ and
$\bar{v}$, respectively~\cite{MITR}. It is gauge equivalent to the
vacuum configuration $\theta_{\mu}(s) = 0$. This configuration gives
rise to a non-zero toron field 
$t_\mu = - (1/L)\,\varepsilon_{\mu\nu} (v_\nu-\bar{v}_\nu)$. The
imaginary part of the effective action is again given by the toron
field contribution. It is non-zero. The real part of the effective
action is found to be divergent as $a_f \rightarrow 0$, in agreement
with the analytic result (\ref{r2}). In Fig.~2 we show 
$\mbox{Re}\,W^\Sigma_- = \mbox{Re}\,W_- + c \sum_{s,\mu} 
2\,(1 - \cos(\theta_\mu(s)))$
as a function of $r$ for two different distances of the vortices. The
divergence is $\propto \log(1/a_f)$. Thus we conclude that these
configurations have zero weight in the partition function. We also
expect that they do not contribute to any observable.

\begin{figure}[tp]
\includemygpt{fig1}{%
  $\mbox{Im} W_-$ as a function of $r$ for the configuration described
  in the text and $(q,\epsilon) = (1,-1)$.}
\end{figure}

\begin{figure}[htbp]
\includemygpt{fig2}{%
$ \mbox{Re} W_-^\Sigma$ as a function of $r$ for the configuration
(11) and $(q,\epsilon) = (1,-1)$.}
\end{figure}

\section{Index theorem}

The lattice action must fulfill the index theorem in order to be in
the same universality class as the continuum action. In general, the 
index theorem
states that the number of zero modes of positive chirality minus the
number of zero modes for negative chirality is equal to the
topological charge, $n_+ - n_- = Q$. In two dimensions it even holds
that $n_+ = Q$ for $Q>0$ and $n_- = |Q|$ for $Q<0$.
Accordingly, a right(left)-handed fermion
has $Q\,\theta(Q)$ ($-Q\,\theta(-Q)$) zero modes.

We have checked that the vector model satisfies the index
theorem. For the eigenvalues we find, both numerically and
analytically, the asymptotic behavior $E_0 = \pi |Q| (a_f/a L)^2$.
For the chiral model, and both gauged and ungauged Wilson terms, the
index theorem is, however, broken. For the ungauged
Wilson term there are no zero modes at all, while for the 
gauged Wilson term we find zero modes of both chiralities like in the
vector model. 

The index theorem can be restored by considering the following Wilson
term
\begin{eqnarray}
S_W^\pm(U^f) &=& \frac{1}{4a_f}\sum_{n,\mu} \bar{\psi}(n) P_\pm 
[2\psi(x)- U^f_{\mu}(n)\psi(n+\mu) - U^{f \,\dagger}_{\mu}(n-\mu)
\psi(n-\mu)] \nonumber \\
             &+& \frac{1}{4a_f}\sum_{n,\mu} \bar{\psi}(n) P_\mp 
[2\psi(x)- \psi(n+\mu) - \psi(n-\mu)]. 
\label{wil}
\end{eqnarray}
The Wilson term (\ref{wil}) has the property 
$S_W^+(U^f) + S_W^-(U^f) = S_W(U^f) + S_W(0)$. We have checked
numerically that it fulfills the index theorem for $|Q| = 1$. 
The next step is to verify the index theorem also for higher 
topological charges.

In the $Q = 0$ sector the Wilson term (\ref{wil}) gives basically the same
results as before. The coefficient $c$ of the counterterm is different
though. In this case we obtain $c = - 0.05971$.

\section{Conclusions}

We may consider the problem of formulating the chiral
Schwinger model on the lattice as being solved for the $Q = 0$
sector. Numerical simulations are now feasible. The real part of the 
effective action is effectively half the action of the corresponding 
vector model, while the imaginary part of the anomaly-free model
can be computed analytically from the toron fields. We hope to be able
to report results on the sectors with non-vanishing topological charge
in the near future.

\section*{Acknowledgements}
One of us (VB) would like to thank DESY-Zeuthen for its hospitality.
This work has been supported in part by INTAS and the Russian
Foundation for Fundamental Sciences through grants INTAS-96-370 and 
96-02-17230a.

\end{document}